\documentclass[a4paper,12pt]{article}

\usepackage{graphicx, color, xcolor, listings, dirtree, authblk, cite, xspace, soul, amsmath, amssymb, url, slashed, multirow, booktabs}  
\usepackage[utf8]{inputenc}	

\usepackage[colorlinks,linkcolor=magenta,citecolor=magenta,urlcolor=magenta]{hyperref}
\usepackage[a4paper,top=2.5cm,bottom=2.5cm,left=2.5cm,right=2.5cm]{geometry}
\usepackage[normalem]{ulem}
\usepackage[flushleft]{threeparttable}
\definecolor{keyword}{HTML}{008000}
\definecolor{emph}{HTML}{0000FF}
\definecolor{string}{HTML}{A52A2A}
\definecolor{comment}{HTML}{004461}
\definecolor{back}{HTML}{F8F8F8}
\definecolor{arrow}{HTML}{745334}

\newcommand{\code}{\texttt}

\newcommand{\es}{\code{EasyScan\_HEP}\xspace}

\newcommand{\footurl}[1]{\footnote{See \url{#1}.}}

\let\oldcite\cite
\renewcommand{\cite}{\unskip~\oldcite}
\newcommand{\reflist}[1]{List.~\ref{#1}}
\newcommand{\reftable}[1]{Tab.~\ref{#1}}
\newcommand{\reffig}[1]{Fig.~\ref{#1}}
\newcommand{\refeq}[1]{Eq.~(\ref{#1})}
\newcommand{\refsec}[1]{Sec.~\ref{#1}}

\newcommand{\fix}[1]{{\color{blue} #1}}

\newcommand{\xnewcommand}[2]{\newcommand{#1}{#2\xspace}}
\newcommand{\latin}[1]{#1}
\xnewcommand{\eg}{\latin{e.g.,}}
\xnewcommand{\ie}{\latin{i.e.,}}
\xnewcommand{\etal}{\latin{et al.}}

\title{\es: a tool for connecting programs to scan the parameter space of physics models}

\author[a,c]{Liangliang Shang \thanks{shangliangliang@htu.edu.cn, first author}}
\author[b,d]{Yang Zhang \thanks{zhangyangphy@zzu.edu.cn, corresponding author}}
\affil[a]{School of Physics, Henan Normal University, Xinxiang 453007, China}
\affil[b]{School of Physics, Zhengzhou University, Zhengzhou 450000, China}
\affil[c]{Department of Physics and Astronomy, Uppsala University, Box 516, SE-751 20 Uppsala, Sweden}
\affil[d]{CAS Key Laboratory of Theoretical Physics, Institute of Theoretical Physics, Chinese Academy of Sciences, Beijing 100190, China}
\date{\today}

\begin{document}


\maketitle

\begin{abstract}

We present an application, \es, for connecting programs to scan the parameter space of High Energy Physics (HEP) models using various sampling algorithms. We develop \es according to the principle of flexibility and usability. \es allows us to connect different programs that calculate physical observables, and apply constraints by one human-readable configuration file. All programs executed through command lines can be connected to \es by setting input and output parameters of the programs. The current version offers the sampling algorithms of Random, Grid, Markov chain Monte Carlo and MultiNest. We also implement features such as resume function, parallelization, post-processing, and quick analysis.

\end{abstract}


\newpage


\noindent
{\bf PROGRAM SUMMARY} 

\bigskip

\begin{small}
\noindent
{\em Program Title: \es}\\[0.5em]
{\em CPC Library link to program files:} (to be added by Technical Editor)\\[0.5em]
{\em Developer's repository link:} \url{https://github.com/phyzhangyang/EasyScan_HEP}\\[0.5em]
{\em Code Ocean capsule:} (to be added by Technical Editor)\\[0.5em]
{\em Licensing provisions:} Apache 2.0\\[0.5em]
{\em Programming language:} Python\\[0.5em]
{\em Supplementary material:} none\\[0.5em]
{\em Nature of problem:} 
Performing numerical analysis of new physics models is crucial in High Energy Physics (HEP), and requires scanning parameter space using various HEP packages. Connecting these packages together can be cumbersome, time-consuming, and prone to errors, especially when using advanced scanning methods, due to the lack of a unified interface between these software and scanning methods.
\\[0.5em]
{\em Solution method:} 
\es utilizes the \texttt{ConfigParser} module in Python to read a unified and human-readable configuration file that connects HEP packages and sets scanning methods. We employ the \texttt{subprocess.Popen} and \texttt{os.system} functions to execute HEP programs, and provide users with various options to configure input parameters and retrieve output parameters, as well as several pre-installed scanning methods.
\\[0.5em]
{\em Additional comments including Restrictions and Unusual features: }
\es is not designed for specific models or HEP packages. Instead, it is compatible with almost any program that can be executed via the command line, requiring minimal interface modifications.
\\[0.5em]

\end{small}

\newpage
\tableofcontents
\newpage

\section{Introduction}
It is necessary to scan the parameter space for numerical analyses in High Energy Physics (HEP).
A variety of HEP packages can be utilized in scanning, which are
spectrum generators~\cite{SPheno,FlexibleSUSY,Eriksson:2009ws,Bauer:2020jay,Allanach:2001kg,Chowdhury:2011zr,Djouadi:2006bz},
observation calculations~\cite{micrOMEGAs,Ge:2016zro,Ge:2020tdh,Guzzi:2014wia,Bringmann:2018lay,Sjostrand:2014zea,GAMBITDarkMatterWorkgroup:2017fax,Coogan:2019qpu,GAMBIT:2017yxo,Gao:2013kp,Denner:2016kdg,Sjodahl:2014opa,Bloor:2021gtp,Ferrando:2010dx,Becher:2011xn,Camarda:2019zyx,Nisius:2014wua,Cullen:2011ac,Athron:2015rva,Hahn:1998yk,Heinemeyer:1998yj,Grazzini:2017mhc,DelDebbio:2013kxa,nCTEQ:2019uwm,Gao:2012ja,Wallraff:2014qka,Stowell:2016jfr,Gauld:2015kvh,Bredenstein:2006rh,Cacciari:2015jma,Fuks:2013vua,Ahrens:2008qu,Borowka:2015mxa,Das:2011dg,Mahmoudi:2009zz,Liebler:2016ceh,Alioli:2008gx,Wainwright:2011kj}, 
event generators or detector simulations~\cite{Forthomme:2018ecc,Dobbs:2001ck,Denner:2014cla,Baranov:2021uol,Harris:2003db,Harland-Lang:2013dia,Buckley:2011ms,Potter:2016pgp,deFavereau:2013fsa,Bellm:2015jjp,Lomnitz:2018juf,Kaspar:2013oka,Kauer:2015hia,Petrov:2007rg,Gehrmann-DeRidder:2014hxk,Ryutin:2017qii,Cacciari:2011ma,Gleisberg:2008ta,Gleisberg:2008fv,Catani:2001cc,Alwall:2011uj,Frixione:2007vw,Corcella:2000bw,Jadach:2015mza,Sjostrand:2019zhc,Dittmaier:2002ap,Ryutin:2011ck,Gingrich:2009da,Denner:2002cg,Delsart:2012jm,Klein:2016yzr,Vakilipourtakalou:2018pfo,Lonnblad:2006pt,Kilian:2007gr,Ellis:2022zdw,Ellis:2020ljj,Ellis:2019zex,Ren:2017jbg,Lu:2015qqa,He:2015spf,He:1998ie,Balazs:1998sb,He:2002fd,He:2002ak}, 
constraint facilities~\cite{HiggsBounds,CheckMATE,Athron:2020sbe,Auffinger:2022dic,GAMBIT:2017qxg,Buckley:2021neu,Buckley:2010ar,Papucci:2014rja,Kluge:2006xs,Bechtle:2013xfa,Camargo-Molina:2013qva,Huang:2016pxg,Bernon:2015hsa,Dorsner:2018ynv,Wiebusch:2012en,IceCube:2012fvn,Kraml:2013mwa,Caron:2016hib,Jeong:2021bpl}, 
and others~\cite{Conte:2012fm,Dai:2007ki,Buehler:2011ev,GAMBITCosmologyWorkgroup:2020htv,Martinez:2017lzg,GAMBITModelsWorkgroup:2017ilg,GAMBITFlavourWorkgroup:2017dbx,Bertone:2014zva,Bertone:2013vaa,Arbey:2011nf,Hou:2019efy,Buckley:2014ana,Sekmen:2018ehb,Carli:2010rw,Buckley:2013jua,Holthausen:2011vd,Li:2022tec,Harz:2016dql,Beneke:2022eci,Binder:2021bmg,Deppisch:2016xlp,Kauer:2001sp,GAMBIT:2018eea,Binoth:2008uq,LHCb:2016mag,Buss:2011mx,Alwall:2006yp,Evans:2016lzo,Przedzinski:2018ett,Hahn:2000kx,Gaunt:2009re,Mertig:1990an,Lind:2020gtc,Andersen:2009nu,Staub:2013tta,Alloul:2013bka,Cacciari:2008gp,Wiebusch:2014qba,Campanario:2013fsa,Salam:2008qg,James:1975dr,Huber:2005yg,Banfi:2012jm,Milhano:2022kzx,Butterworth:1996zw,Klappert:2020nbg,Butterworth:2002xg,Feger:2019tvk,Ilakovac:2014yka,Rubin:2010xp,Parkes:2016yie,Czakon:2005rk,Gao:2013bia,Dasgupta:2014yra,Polesello:2009rn,Poluektov:2014rxa,Karamitros:2021nxi,Savvidy:2014ana,Bella:2010vi,Monk:2014uza,Forte:2002fg,Peraro:2014cba,Evslin:2015pya,Cobanoglu:2010ie,Gehrmann:2018yef,Becher:2011fc,DELPHI:1996sen,Chekanov:2013mma,Botje:2010ay,Beneke:2016kkb,Antusch:2005gp,Actis:2012qn,vonManteuffel:2012np,Feng:2012iq,Falkowski:2015wza,Guzzi:2021gvv,Mastrolia:2010nb,Coimbra:2013qq,Harland-Lang:2019eai,Salam:2007xv,Bell:2018vaa,Cardoso:2015gfa,Murakami:2013rca,Maurer:2015gva,Alver:2008aq,Goodsell:2019zfs,Abdulov:2021ivr,Almeida:2010pa,Bzowski:2013sza,Gao:2010qx}, 
for instance. 
We should take an assortment of time and effort on data transmission by implementing scripts when a large number of packages are employed.
In addition, connections between packages become more complicated 
when sophisticated sampling methods such as Markov Chain Monte Carlo (MCMC)~\cite{mcmc1,mcmc2} and MultiNest~\cite{MultiNest} are adopted.

As a result, we develop a user-friendly program named \es to make scanning easier and more efficient.
\es can easily connect external packages, which is different to the sophisticated scan programs like \code{GAMBIT}~\cite{GAMBIT}, \code{MasterCode}~\cite{MasterCode}, \code{SuperBayeS}~\cite{SuperBayeS} and \code{Fittino}~\cite{Fittino} that require modifying the interface of external HEP packages~\footnote{\code{xBit}~\cite{Staub:2019xhl} and \code{BSMArt}~\cite{Goodsell:2023iac} are designed with a similar purpose to \es, but they have different focuses and usages.}. To achieve this, \es provides several flexible methods for setting input and output parameters.

In detail, \es can rescue us from tedious and inefficient scanning in the following aspects,
\begin{itemize}
    \item Connection of external packages. Users only need to provide information about input and output parameters, execution commands, and the locations of external packages. \es will automatically call these packages during scanning.
    \item Variety of sampling algorithms. It is convenient to migrate one sampling algorithm to another one in \es because different sampling algorithms own one common interface. The sampling algorithms of OnePoint, Random, Grid, MCMC and MultiNest are provided now and more algorithms will be provided in the future, such as the ongoing program \code{GAMBIT\_light} and machine learning driven sampling methods~\cite{Ren:2017ymm,Hammad:2022wpq}. 
    \item Analyses of results. Once the scanning is complete, \es immediately provides features of results such as histograms, scatter plots, and contour plots.
\end{itemize}
Users can configure the above settings in a human-readable configuration file. All results are saved in an organized manner within a folder. Additionally, we offer some useful features, such as the ability to resume from a break-point and post-processing.

During the development of \es, it has been used in several works~\cite{Pozzo:2018anw,Athron:2017drj,Cao:2018iyk,Cao:2017ydw,Cao:2017sju,Cao:2017cjf,Shang:2019zhh, Shang:2020clm, Yang:2019uea, Yang:2021btv, Shang:2021mgn, Wang:2021bcx, Yang:2021dtc, Shang:2022hbv, Yang:2022gvz, Yang:2022wfa, Shang:2022tkr, Shang:2023rfv,Wang:2023suf}. 
The configuration file of Ref.~\cite{Yang:2022gvz} is provided as an example of \es 
with \code{makefile} to install the relevant HEP packages, 
which is shown in \refsec{sec:example}. From this aspect, \es can help people to repeat the works that use \es to organize scanning.

The outline of our paper is as follows. We give a quick start in \refsec{sec:install} including installation and a toy example. 
Then we describe the structure of \es in \refsec{sec:structure}.
Definitions and conventions of items in the configuration file are given in \refsec{sec:configuration}. \refsec{sec:example} provides a practical example configuration file. The summary is given in \refsec{sec:summary}.


\section{Quick start}\label{sec:install}

\begin{figure}[ht]
\centering
\includegraphics[width=0.37\textwidth]{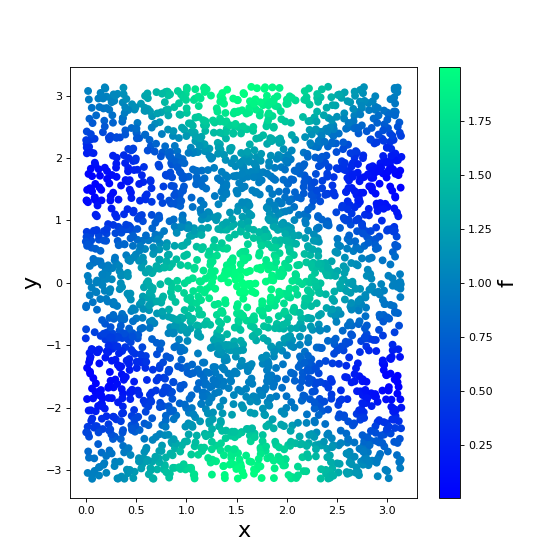}
\includegraphics[width=0.37\textwidth]{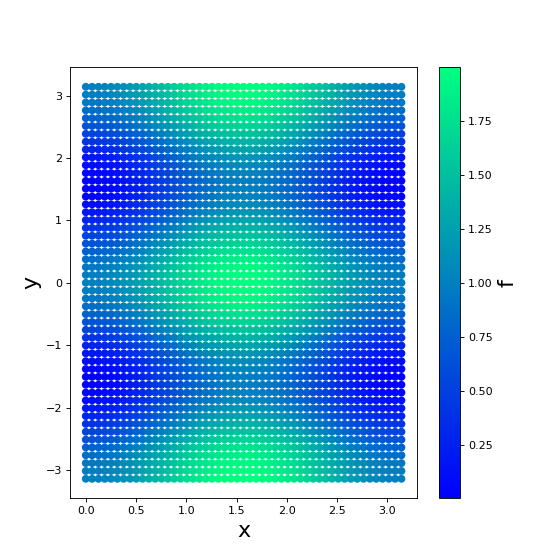}
\includegraphics[width=0.37\textwidth]{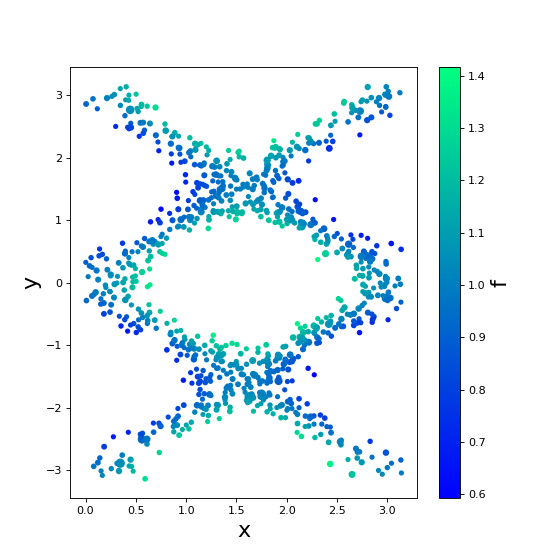}
\includegraphics[width=0.37\textwidth]{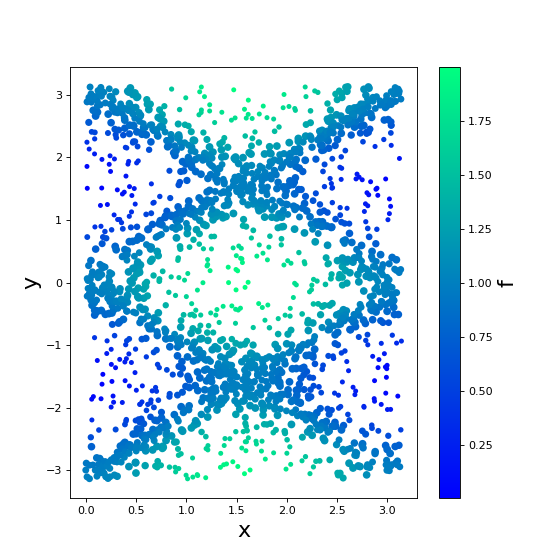}
\caption{Four output plots corresponding to the configuration files of \es, \code{example\_random.ini} (upper left), \code{example\_grid.ini} (upper right),  \code{example\_mcmc.ini} (lower left) and  \code{example\_multinest.ini} (lower right). Each plot contains approximately 2500 samples.}
\label{fig:test_run}
\end{figure}

\es is written in \code{Python3} with dependencies on \code{numpy}\cite{numpy}, \code{scipy}\cite{scipy} and \code{ConfigParser} libraries.
An optional plot function and the MultiNest sampler further require \code{matplotlib}\cite{matplotlib}, \code{pandas}\cite{pandas} and \code{pymultinest} \cite{Buchner:2014nha} libraries, respectively. These libraries can be installed via \code{pip}:
\begin{lstlisting}
$ sudo apt install python3-pip python3-tk
$ pip3 install numpy scipy matplotlib ConfigParser pandas pymultinest
\end{lstlisting}

The repository for \es is available at 
\begin{center}
  \url{https://github.com/phyzhangyang/EasyScan_HEP}.
\end{center}
To download \es, the following commands can be utilized:
\begin{lstlisting}
$ git clone https://github.com/phyzhangyang/EasyScan_HEP.git
\end{lstlisting}

\code{easyscan.py} in the folder named \code{bin} is the main program file, which can be executed with a configuration file by the command, for example, 
\begin{lstlisting}
$ ./bin/easyscan.py templates/example_random.ini
\end{lstlisting}
Besides the above normal mode, \es can be executed in debug mode by
\begin{lstlisting}
$ ./bin/easyscan.py templates/example_random.ini -d
\end{lstlisting}
with more screen outputs for debugging, in resume mode by 
\begin{lstlisting}
$ ./bin/easyscan.py templates/example_random.ini -r
\end{lstlisting}
in case a previous scan is accidentally aborted, and in parallel mode by modifying \code{Parallel threads} in \code{example\_random.ini} larger than one.
Here \code{example\_random.ini} is an example configuration files provided in \es. 
The command performs a random scan based on a simple function,
\begin{equation}\label{eq:testmodel}
f(x,y) = \sin^2 x + \cos^2 y, 
\end{equation}
where $x$ and $y$ are two input parameters in the range of $[0,\pi]$ and $[-\pi,\pi]$, respectively, and $f$ is an output parameter. 
If the example runs successfully, a folder named \code{example\_random} will be generated, and it will contain an image named \code{test\_run\_random.png} in the \code{example\_random/Figures} folder. This image is similar to the upper left panel of \reffig{fig:test_run} but with fewer data points.

In the \code{templates} folder of \es, we provide three other instructive configuration files, i.e., \code{example\_grid.ini}, \code{example\_mcmc.ini} and \code{example\_multinest.ini}.
These files perform scanning based on the same function (Eq.~\ref{eq:testmodel}) but with different samplers compared to the above example. 
They will generate plots similar to the remaining panels of \reffig{fig:test_run}. 
What is more, the likelihood function in \code{example\_mcmc.ini} and \code{example\_multinest.ini} is
\begin{equation}\label{eq:testlike}
\mathcal{L}=\frac{1}{\sqrt{2\pi\sigma^2}}\exp\left(\frac{-(f-\mu)^2}{2\sigma^2}\right),
\end{equation}
where the mean $\mu=1$ and the standard deviation $\sigma=0.2$ are set.

Now, we will explain the configuration file \code{example\_random.ini} to provide a brief demonstration of \es usage. All settings are presented in a human-readable format. For more detailed instructions on how to use the configuration file, users can refer to \refsec{sec:configuration}.

\begin{lstlisting}[backgroundcolor=\color{back},frame=trBL,numbers=left,numberstyle=\itshape\color{comment},xleftmargin=2.5em,framexleftmargin=2em,caption={Contents of the configuration file \code{example\_random.ini}}, label={list:example}]
[scan]
Result folder name: example_random
Sampler:            random
#                   VarID  Prior   Min      MAX 
Input parameters:   x,     flat,   0,       3.14
                    y,     flat,   -3.14,   3.14
Points number:      100
Parallel threads:   1
Parallel folder:    utils

[program1]
Execute command:   ./TestFunction.py
Command path:      utils/
#                  FileID    FileName
Input file:        1,        utils/TestFunction_input.dat
#                  VarID     FileID  Method    Row Column
Input variable:    x,        1,      Position, 1,  1
                   y,        1,      Position, 1,  2
Output file:       1,        utils/TestFunction_output.dat
#                  VarID     FileID  Method    Row Column
Output variable:   f,        1,      Position, 1,  2

[constraint]
#                  VarID     Mean   StandardDeviation
Gaussian:          f,        1,     0.2

[plot]
#                  X-axis    Y-axis     Color   FigureName
Color:             x,        y,         f,       test_run_random
                   x,        y,         Chi2
\end{lstlisting}

\reflist{list:example} displays the entire content of the configuration file \code{example\_random.ini}. This file is divided into four sections, each of which contains keys and values following the \code{ConfigParser} format\footnote{See \url{https://docs.python.org/3/library/configparser.html} for the rules of \code{ConfigParser}.}. 
\begin{description}
\item[{[scan]}] (\code{line~1 $\sim$ line~9}) This section is used to configure certain sampling parameters and includes the following keys:
    \begin{itemize}
        \item \code{Result folder name}: It specifies the name of the folder where all results of \es are saved. The value for this key in the example is \code{example\_random}. When \es is initiated, a folder named \code{example\_random} will be created.
        \item \code{Sampler}: It is used to choose a sampler. In this case, random scanning is utilized.
        \item \code{Input parameters}: It defines all input parameters for the scanning process, with each line representing a parameter description. For example, Line~5 in the example corresponds to the parameter $x$ in \refeq{eq:testmodel}, and Line~6 represents $y$ in \refeq{eq:testmodel}. Each line consists of four arguments separated by commas: the parameter name, the prior distribution, the lower limit, and the upper limit of the parameter.
        \item \code{Points number}: It specifies the number of samples, and its interpretation varies for different samplers, as explained in \refsec{sec:scan}.
    \end{itemize}

\item[{[program1]}] (\code{line 11 $\sim$ line~21}) This section is used to control external programs, also referred to as HEP packages, responsible for calculating observables or other results.
    
In the example above, we utilize \code{utils/TestFunction.py} to simulate an external program responsible for calculating $f(x, y)$ as mentioned. It is executed using the command:
\begin{lstlisting}
$ ./TestFunction.py
\end{lstlisting}
The input file for \code{TestFunction.py} is \code{TestFunction\_input.dat}, where the first floating number represents $x$, and the second floating number represents $y$. For example,
\begin{lstlisting}[backgroundcolor=\color{back},frame=trBL,caption={Contents of the input file \code{`TestFunction\_input.dat'} }, label={list:example_input}]
1.5845887764980207	2.95977735836697
\end{lstlisting}
The output file for \code{TestFunction.py} is \code{TestFunction\_output.dat}, which contains a single floating number $f$. For example,
\begin{lstlisting}[backgroundcolor=\color{back},frame=trBL,caption={Contents of the output file     \code{TestFunction\_output.dat} }, label={list:example_output}]
f= 1.96711562785
\end{lstlisting}
    
Therefore, \code{[program1]} includes the following items in order to use \code{TestFunction.py}:
    \begin{itemize}
        \item \code{Execute command}: It defines the command for executing HEP packages. In this case, its value is \code{./TestFunction.py}.
        \item \code{Command path}: It specifies a relative path to the location where \es is executed or an absolute path, and it is used as the reference point for running the commands provided in \code{Execute command}. In this case, a relative path is used.
        \item \code{Input file}: It provides the location and name of certain input files. The first element of its value is an identification number (\code{ID}), which is used to identify the input file. In \reflist{list:example}, it is specified as \code{ID=1}. The second element is the relative or absolute path and name of the input file.
        \item \code{Input variable}: It configures input parameters, and each line takes one parameter. The first element in each line is the name of the parameter, the second element is the \code{ID} of the file in which the parameter is located, and the remaining elements specify how to locate the parameter. For instance, Line~15 in \reflist{list:example} indicates that $x$ is situated at the first row and first column in the input file with \code{ID=1}. In addition to the \code{Position} method, there are several other methods described in \refsec{sec:program}.
        \item \code{Output file}: It specifies the location and name of certain output files, following rules similar to those of \code{Input file}.
        \item \code{Output variable}: It explains how to retrieve output variables, with rules akin to those of \code{Input variable}, except for the \code{Replace} method. In this example, Line~19 in \reflist{list:example} signifies that $f$ obtains its value from the first row and second column in the output file with \code{ID=1}.
    \end{itemize}
\item[{[constraint]}] (\code{line 23 $\sim$ line~25}) This section establishes constraints that are applied during the scanning process, including:
    \begin{itemize}
        \item \code{Gaussian}: It establishes a constraint of a one-dimensional Gaussian distribution. The likelihood function, as defined in \refeq{eq:testlike}, is implemented by setting the variable to $f$, the mean to $1$, and the standard deviation to $0.2$. The resulting Gaussian chi-square value will be saved in an output file named \code{Gaussian\_f}.
        This constraint is not utilized to guide random and grid scans but is crucial for more advanced samplers, such as MCMC and MultiNest.
        \es can calculate the confidence based on the total $\chi^2$ to determine whether a sample is excluded or not.
    \end{itemize}
\item[{[plot]}] (\code{line 27 $\sim$ line~30}) This section generates plots after the completion of all sampling processes.
    \begin{itemize}
        \item \code{Color}: It is used to create colorful scatter plots. Each line takes care one picture. The first element in a line represents the X-axis, the second element represents the Y-axis, the third element represents the color, and the last element represents the graph name. An image named \code{test\_run\_random.png} will be generated in the folder \code{example\_random/Figures}.
    \end{itemize}
\end{description}

Upon completion of the scanning process, all results will be stored in the folder named \code{example\_random}. For random scanning in the example, it includes:
\begin{itemize}
    \item \code{ScanResult.txt}. It is used to save all numerical results when \es successfully completes the scan. Each line corresponds to one sample, excluding the first row. The first row serves as a header that explains the meaning of each column. The subsequent rows contain numerical results, including all variables from \code{Input variable} and \code{Output variable} in each \code{[program\#]} section \footnote{\code{\#} represents a number.}, as well as \code{Gaussian} and \code{FreeFormChi2} from the \code{[constraint]} section. Additionally, there is an extra column called \code{Chi2}, indicating the total $\chi^2$. We provide the first three rows from \code{ScanResult.txt} in \reflist{list:example_result}.
\begin{lstlisting}[backgroundcolor=\color{back},frame=trBL,numbers=left,numberstyle=\itshape\color{comment},xleftmargin=2.5em,framexleftmargin=2em,caption={Some contents of \code{ScanResult.txt} in the example in \refsec{sec:install}},  label={list:example_result}]
 x,     y,      f,     Chi2,  Gaussian_f
 2.18,  -0.99,  0.96,  0.02,  0.02
 2.23,  -0.27,  1.54,  7.39,  7.39
\end{lstlisting}      
    \item \code{Figures/}. It is used to store the figures defined in the \code{[plot]} section.
    \item \code{SavedFile/}. It is used to save the output files defined in the \code{[program\#]} sections, but it is empty in this case.
\end{itemize}
The contents of the output folders for the other samplers are explained in \refsec{sec:scan}. For additional guidelines on the rules pertaining to a configuration file, users can refer to \refsec{sec:configuration}.

\section{Structure}\label{sec:structure}

\begin{figure}[h!]
\centering
\includegraphics[width=\textwidth]{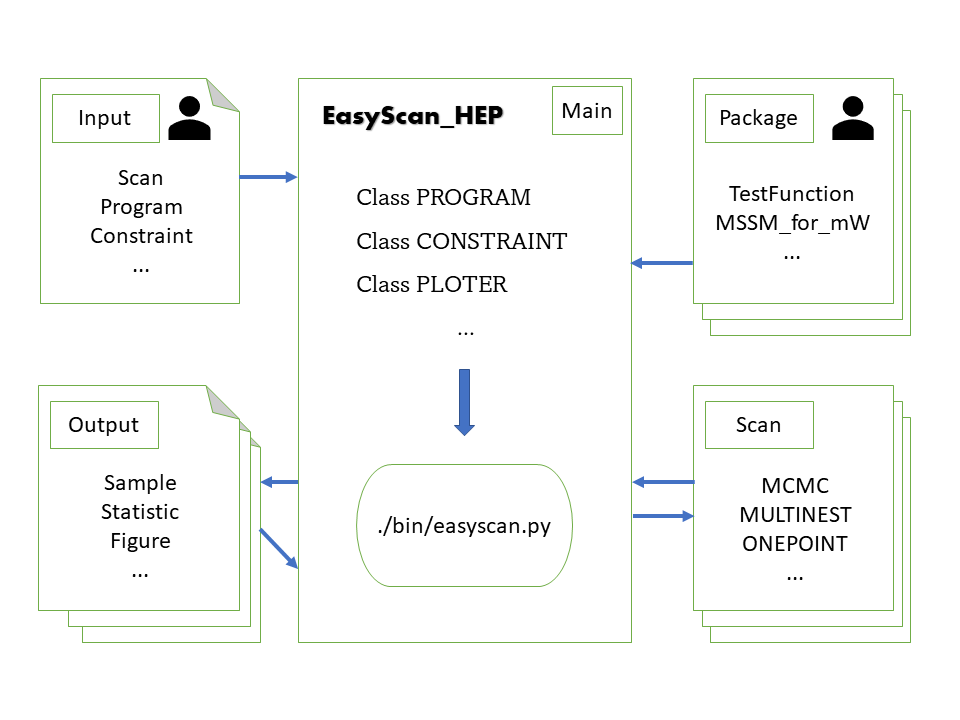}
\caption{Workflow of \es.}
\label{fig:workflow}
\end{figure}

The structure of \es$\,$ is depicted in \reffig{fig:workflow}, which consists of five blocks. The \code{Input} block necessitates information regarding the sections, \code{[scan]}, \code{[program\#]}, \code{[constraint]}, and \code{[plot]}, as explained in \refsec{sec:install}. The details from \code{[scan]} determines the sampler used in the \code{Scan} block. The \code{Scan} block forwards the sampling data to the \code{Main} block, which subsequently puts this data into input files of each HEP package in the \code{Package} block. The \code{[Main]} block invokes programs in the numerical order. If a sophisticated sampler, such as MCMC or MultiNest, is chosen, the \code{Scan} block is also responsible for determining the scanning direction. 
During \es running, it saves variables into files, but the generation of figures occurs only after the entire sampling process is completed.

\begin{figure}[h!]
\centering
\framebox[\textwidth]{%
\begin{minipage}{0.95\textwidth}
\vspace{2mm}
	\dirtree{%
		.1 \es.
		.2 bin/\DTcomment{Executable file}.
		.3 easyscan.py.
		.2 src/\DTcomment{Internal functions}.
		.3 easyscan\_logging.conf.
		.3 program.py.
		.3 initialize.py.
		.3 readin\_config.py.
        .3 scanner.py.
        .3 scan\_controller.py.
        .3 constraint.py.
	    .3 statfun.py.
	   	.3 ploter.py.
	    .3 auxfun.py.
		.2 utils/\DTcomment{Auxiliary functions}.
		.3 TestFunction.py.
		.3 TestFunction\_input.dat.
        .3 OnePointBatch.in.
		.3 MSSM\_mW.
		.2 templates/\DTcomment{Example configuration files}.
	    .3 example\_random.ini.
	    .3 example\_grid.ini.
	    .3 example\_mcmc.ini.
	    .3 example\_mcmc\_bound.ini.
	    .3 example\_multinest.ini.
	    .3 example\_onepoint.ini.
	    .3 example\_onepointbatch.ini.
	    .3 example\_plot.ini.
	    .3 scan\_MSSM\_for\_mW.ini.
	    .3 bound.txt.
		.2 README.rst \DTcomment{Readme}.
		.2 LICENSE \DTcomment{Apache license}.
	}
\vspace{2mm}
\end{minipage}
}
\caption{Directory structure of \es.}\label{fig:DirectoryTree}
\end{figure}

The directory structure of \es is illustrated in \reffig{fig:DirectoryTree}, and its meaning is explained as follows:
\begin{itemize}
\item \code{bin}: The primary executable program, named \code{easyscan.py}, is situated in this directory.

\item \code{src}: The internal functions of \es are located in this directory. \code{initial.py} is utilized for initializing the parameters of \es, whereas \code{program.py} serves to connect the programs defined in the \code{Package} block, establish input variables, and extract output variables from output files.
\code{scanner.py} is responsible for conducting sampling based on various samplers. \code{statfun.py} offers prior probabilities, and \code{scan\_controller.py} provides additional functions for scanning. \code{constraint.py} is employed for calculating Gaussian and free-form chi-squares. \code{readin\_config.py} is used to parse a configuration file, and \code{ploter.py} is responsible for generating graphs. Lastly, \code{auxfun.py} supplies common functions used in \es.

\item \code{utils}: HEP packages, such as \code{SPheno} and \code{Micromegas}, are typically found in this directory. However, it's important to note that these packages can be located anywhere, as long as their paths are defined in a configuration file, as explained in \refsec{sec:install}.

\item \code{templates}: This directory contains various configuration files for different samplers that users can reference.

\end{itemize}

\section{Configuration file} \label{sec:configuration}

The \code{ConfigParser} module in Python is employed to parse the configuration file, and it organizes the file into sections, followed by keys and their corresponding values. Here are some general rules:
\begin{itemize}
    \item The header of each section is identified by a line with the format \code{[section\_name]}, like \code{[scan]}. Section names are case-sensitive.
    \item All sections in \es are mandatory, except for the \code{[plot]} section if quick analyses of results are not required, and the \code{[constraint]} section if samplers do not need likelihood. Additionally, the number in \code{[program\#]} (\# representing for a number) is arbitrary.
    \item In each section of \es, a key is denoted by items before a colon symbol. Keys represent functions and should not be altered. The values, located after the colon symbol, come in various formats depending on the specific key.
    \item The keys within each section are fixed and not case-sensitive. The required keys for each section may differ depending on the specific sampler, and any unnecessary keys will be disregarded.
    \item The initial set of values must be on the same line as their key, while subsequent sets of values should be positioned on new lines.
    \item Lines that start with a \code{\#} character are regarded as comments and will be disregarded by \es.
\end{itemize}
In the subsequent subsections, we will outline the guidelines for formatting values for all sections and their respective keys.

\subsection{\code{[scan]}} \label{sec:scan}

\begin{table}[ht]
\caption{Rules for the \code{[scan]} section}
\centering 
\begin{threeparttable}
\begin{tabular}{|c|c|c|c|c|}
\hline
Key                & Type of value                          & Mandatory & Multi-line & \# of component  \\\hline
Result folder name & string                                 & Yes       & No        & 1          \\\hline
Sampler            & \begin{tabular}[c]{@{}c@{}}
                       \textbf{OnePoint} \\
                        \textbf{Random} \\ 
                        \textbf{Grid}   \\ 
                        \textbf{MCMC}   \\ 
                        \textbf{MultiNest}\\
                        \textbf{Plot} \\
                        \textbf{Postprocess}
                     \end{tabular}                          & Yes       & No        & 1         \\\hline
Input parameters   & \begin{tabular}[c]{@{}c@{}}
                        string,\textbf{Flat},float,float,$*$ \\ 
                        string,\textbf{Log},float,float, $*$ \\
                        string,\textbf{Fixed},float 
                     \end{tabular}                          & Yes       & Yes       & $\geq3$   \\\hline
Points number   & integer                                & Yes/No
                                                      \tnote{$\dagger$} & No        & 1         \\\hline
\end{tabular}\label{tab:scan1}
  \begin{tablenotes}
  \item[$\dagger$] No for the \code{Grid} sampler.
  \end{tablenotes}
\end{threeparttable}
\end{table}

This section is utilized to define the fundamental scanning parameters. Each piece of information consists of keys and values, which can vary based on the selected sampler.
\begin{itemize}
\item \code{Result folder name}

This key specifies the name of the folder where \es results are stored. The folder is created in the directory where \code{easyscan.py} is executed. If a folder with the same name already exists before running \code{easyscan.py}, \es will present the following message:
\begin{lstlisting}
$ * The Result folder [test] already exists.
$ Choose: (r)replace , (b)backup , (s)stop
\end{lstlisting}
where `r' signifies deleting the old folder and creating a new one, `b' represents creating a backup in a folder named \code{Backup}, which is located in the same path as the original folder, and then creating a new one. `s' indicates stopping the operation and exiting.

\item \code{Scan method}

The value for \code{Scan method} must be selected from the following four options, and the choice is not case-sensitive:

\begin{itemize}
\item \code{Random}: This option is employed for scanning the parameter space in a random manner.
\item \code{Grid}: This option is utilized for conducting a grid scan in the parameter space, where equidistant intervals are defined for each parameter.
\item \code{MCMC}: This option is based on the Metropolis-Hastings algorithm~\cite{mcmc1}.
\item \code{MultiNest}: This option is based on the MultiNest algorithm~\cite{MultiNest}.
\item \code{OnePoint}: This option is employed to compute a single sample, usually for testing a configuration file. If there is an incorrect setting, \es will generate warnings. An absolute path to a data file can also be specified after the \code{OnePoint/} symbol, and \es will treat each line as one data point. This is the batch mode of \code{OnePoint}. The file should include a header in the first line, which should correspond to the names of all scanning parameters.
\item \code{Plot}: This option generates certain figures using the parameters acquired previously. To utilize this mode, an output file, such as \code{ScanResult.txt}, should already be available. The other settings in \code{[scan]}, \code{[program]}, and \code{[constraint]} are disregarded.
\item \code{Postprocess}: This option is utilized to post-process the scan previously conducted by \es, such as incorporating new observables and recalculating certain observables.
\end{itemize}

\item \code{Input parameters}:
Values for \code{Input parameters} can be entered on multiple lines, with each line representing one parameter. Parameter names should start with a letter, be unique compared to other parameters, and are case-sensitive. Prior distributions for input parameters can be chosen from \code{Flat}, \code{Log}, and \code{Fixed}. Each line should adhere to the following rules, depending on the selected sampler:

\begin{itemize}
  \item For the \code{Random} sampler, each line should be structured in the following order: parameter name, prior distribution, minimum value, and maximum value. These four items, collectively referred to as Basic-Items, should be separated by commas. For example:
      \begin{lstlisting}
#                 VarID, Distribution, Min, Max
Input parameters: tanb, Flat, 2.0, 60.0
                   m0,  Log,  500, 5000
      \end{lstlisting}
  \item For the \code{Grid} sampler, the items in a line should include the Basic-Items, followed by the number of intervals. If the number of intervals is not specified, the default value of 10 will be applied. For example, if the number of intervals in the above example is set to 10, 11 points will be generated as follows: $\tan\beta=2.0,7.8,13.6,...,60.0$. An example is provided below:
      \begin{lstlisting}
#                 VarID, Distribution, Min, Max, NumberOfIntervals
Input parameters: tanb, Flat, 2.0, 60.0, 10
      \end{lstlisting}
  \item For the \code{Multinest} sampler, only Basic-Items are employed, and they are identical to those used for the \code{Random} sampler.
  \item For the \code{MCMC} sampler, each line should include the Basic-Items, the number of intervals for each parameter, and its initial value. If the initial value is not specified, the default value will be set as the average of the minimum and maximum values of the parameter.
  For instance, $\tan\beta=8$ will be used as its initial value in the example below:
      \begin{lstlisting}
#          VarID, Distribution, Min, Max, NumberOfIntervals, Initial
Input parameters: tanb, Flat, 2.0, 60.0, 10, 8.0
      \end{lstlisting}
  \item For the \code{OnePoint} sampler, each line should include the name, \code{Fixed}, and value of each parameter in sequence. However, in the batch mode of \code{OnePoint}, the value of the parameter can be omitted.
\end{itemize}

   \item \code{Points number}: The value for \code{Points number} should be a positive integer, indicating the number of randomly generated points for the \code{Random} sampler, the number of reserved samples for the \code{MCMC} sampler, and the number of live samples in the ensemble for the \code{Multinest} sampler. If this value is not provided, the default value of 10 will be applied. However, for the \code{OnePoint} and \code{Grid} samplers, this field is unnecessary and can be omitted.

   \item \code{Interval of print}: The \code{Interval of print} should be a positive integer, and it signifies the frequency at which \es outputs are displayed. For example, if the value is one, the output for each scanned point (or reserved point for \code{MCMC}) will be displayed. If the value is two, the output for the `1st, 3rd, 5th, ...' scanned point (or reserved point for \code{MCMC}) will be displayed. If this value is not provided, the default value of one will be applied.
   
   \item \code{Random seed}: The \code{Random seed} should be a real number, and it serves as the random number seed for the \code{Random} and \code{MCMC} samplers. If this value is not specified, the random number seed will be set based on the system time when \es starts running.

   \item \code{Parallel threads}: This gives the number of threads will be used in parallel mode, which has to be a positive integer. 
   In \reftable{tab:time}, we show the elapsed time for \code{example\_random\_parallel.ini} with varying numbers of parallel threads, to demonstrate the time that parallelization can save. These measurements were taken on a laptop equipped with an Intel Core i7-12800H processor. To simulate real-life scenarios, a ``\code{sleep(0.5)}" function call has been added to ``\code{tools/TestFunction.py}".

\begin{table}[ht]
\caption{Elapsed time comparison for \code{example\_random\_parallel.ini} with different numbers of parallel threads. 
}
\centering 
\begin{threeparttable}
\begin{tabular}{|c|c|c|c|c|c|c|c|c|c|c|c|c|c|c|}
\hline
Num of threads  & 1   & 2  & 3   & 4   & 5   & 6  & 7  & 8   & 9   & 10  & 11  & 12  & 13  & 14 \\\hline
Time (seconds)  & 58  & 29 & 20  & 15  & 13  & 11 & 10 & 9.1 & 8.4 & 7.4 & 7.1 & 6.9 & 6.5 & 6.0 \\\hline
\end{tabular}\label{tab:time}
\end{threeparttable}
\end{table}

   \item \code{Parallel folder}: This folder contains all the external programs that are called during the scanning process. To avoid modifying the external program, we parallelize the scan by creating copies of this folder. We then run the external programs in each separate folder on a single thread. It is essential to ensure that the external program can execute successfully and independently in each copied folder. 

   \item[*] In parallel mode, the \code{Command path}, \code{Input file} and \code{Output file} of each \code{[program\#]} described in the next subsection must start with the  \code{Parallel folder}.

   \item[*]In parallel mode, the value for \code{Points number} is the total number of samples in the \code{Random} sampler, the total number of reserved samples for the \code{MCMC} sampler, and the total number of live samples in the ensemble for the \code{Multinest} sampler.
   
   
\end{itemize}

\subsection{\code{[program\#]}} \label{sec:program}

\begin{lstlisting}[ caption={An example of some \code{[program\#]} sections.}, captionpos=b]
[program1]
Program name:      softsusy-3.7.3
Execute command:   ./softsusy-3.7.3/softpoint.x leshouches < softsusy-3.7.3/inOutFiles/lesHouchesInput > LesHouchesOutput.txt
Command path:      utils/
Input file:        1, utils/softsusy-3.7.3/inOutFiles/LesHouchesInput.txt
#                  VarID FileID   Method     others
Input variable:    m0,    1,   Position,  14, 2
                   m12,   1,   Label,     m12, 2
                   tanb,  1,   Replace,   ES_tanb
                   A0,    1,   SLHA,      BLOCK, MINPAR, 5
Output file:       1, utils/lesHouchesOutput
#                  VarID FileID Method others
Output variable:   mh,    1,   SLHA,  BLOCK, MASS, 25
                   mn1,   1,   SLHA,  BLOCK, MASS, 1000022
                   spec,  1,   File,  SAVE
#                  VarID  LowLimit UpLimit
Bound:             mh,    120,     130

[program2]
Program name:      micromegas_4.3.1
Execute command:   ./micromegas_4.3.1/MSSM/main LesHouchesOutput.txt > micromrgas.out
Command path:      utils/
Input file:
Input variable:
Output file:       1, utils/micromrgas.out
#                  VarID  FileID  Method   others
Output variable:   omega,   1,   label, Omega=, 3
Time limit minute: 10           
\end{lstlisting}

This section furnishes information about the HEP packages used in the scanning process. Each HEP package is assigned its own section, and the section name should adhere to the format of \code{[program\#]}, where \code{\#} is a positive integer that must be unique for each section. The sections will execute in numerical order based on \code{\#} rather than the order in which they are written. The keys for each section are as follows:
\begin{itemize}
  \item \code{Program name}: The \code{Program name} values are strings that serve as identifiers for a HEP package.
  \item \code{Execute command}: The \code{Execute command} values are utilized to execute HEP packages and can be written across multiple lines, with each line representing a single command. It's important to note that these values should not contain any commas since commas are interpreted as separators.
  \item \code{Command path}: The \code{Command path} values specify the location where the commands provided in \code{Execute command} should be executed. These values should either be an absolute path or a path relative to the directory where \code{./easyscan.py} is executed. It is recommended to use an absolute path for clarity and reliability.
  \item \code{Input file}: The \code{Input file} values can span across multiple lines, with each line representing a single input file. Each line should include two elements: an \code{ID} that uniquely identifies the input file and the file's path and name. The path can be specified as either an absolute path or a path relative to the directory indicated in \code{Command path}. Elements within a line should be separated by commas, and it is advisable to use absolute paths instead of relative paths for clarity and consistency.
\begin{lstlisting}[ caption={Some contents of an example input file.}, captionpos=b, label={lstfile}]
  	Block MINPAR
  	1 1.250000000e+02   # m0
  	2 9.000000000e+02   # m12
  	3 1.000000000e+01   # tan beta at MZ
  	4 1.000000000e+00   # sign(mu)
  	5 0.000000000e+00   # A0
  	BLOCK UMIX # Chargino Mixing Matrix U
  	1 1 -9.69283723E-01 # U_11
  	1 2 2.45945247E-01  # U_12
  	2 1 2.45945247E-01  # U_21
  	2 2 9.69283723E-01  # U_22
  	DECAY 1000023 1.22776237E-04 # neutralino2
  	8.88469593E-04 2 1000022 22   #BR(~ chi_20 -> ~chi_10 gam)
  	1.15050111E-01 3 1000022 -2 2 #BR(~ chi_20 -> ~chi_10 ub u)
\end{lstlisting}
  
\begin{lstlisting}[ caption={Some example contents of an input file for the \code{Replace} method. }, captionpos=b, label={lst1}]
Block MINPAR # Input parameters
	1 ES_m0             # m0
\end{lstlisting}
  
  \item \code{Input variable}: The \code{Input variable} values can span across multiple lines, and each line represents a single input parameter. Input parameters can be sourced from either the \code{Input parameters} in the \code{[scan]} section or the \code{Output variable} of a previous \code{[Program\#]} section. Each line should consist of three parts: the parameter name, the ID of the input file where the parameter is located, and additional elements that specify how to organize the parameter. There are four available methods: \code{Position}, \code{Replace}, \code{Label}, and \code{SLHA}. Here are examples of how to configure \code{m0} in the \code{ID=1} input file shown in \reflist{lstfile} using the input parameter named `m0':
  \begin{itemize}
     \item \code{Position}: Since the value of \code{m0} is located in the second row and second column of the input file (\reflist{lstfile}), where columns are separated by tabs or spaces, it can be configured using the input parameter `m0' with the \code{Position} method as follows:
\begin{lstlisting}
#              VarID, FileID, Method,  Row, Column
Input variable: m0, 1, Position, 2, 2
\end{lstlisting}
    
     \item \code{Label}: If the value is situated in a line containing a unique symbol, such as \code{m0}, which appears only once in the \code{ID=1} input file, and the value is the second item in the line with items separated by tabs or spaces, you can configure it using the input parameter `m0' with the \code{Label} method as follows:
\begin{lstlisting}
#              VarID FileID Method Identifier Column
Input variable:  m0,   1,  Label,  m0,     2
\end{lstlisting}
The input parameter will be assigned to the second item of the line identified by the \code{m0} symbol. It's important to note that the numbering of items in a line begins from 1 for the first item, 2 for the second item, and so forth. 

     \item \code{SLHA}: Given that the value of \code{m0} is situated in the \code{MINPAR} block with an index of \code{1} in an SLHA format file~\cite{Skands:2003cj}, users can configure it using the \code{SLHA} method as follows:
\begin{lstlisting}
#               VarID FileID Method Section SectionName Keys
Input varialbe: m0,  1,  SLHA, BLOCK,  MINPAR,  1
\end{lstlisting}
The input file also contains other structures, such as \code{UMIX} and \code{DECAY}, which can be handled as follows:
\begin{lstlisting}
#               VarID      FileID Method Section SectionName Keys
Input variable: U11,       2, SLHA, BLOCK, UMIX, 1, 1
	        Gamma_n2,  2, SLHA, DECAY, 1000023, 0
	        BRn2_n1Ga, 2, SLHA, DECAY, 1000023, 2, 1000022, 22
\end{lstlisting}       
In the above examples, \code{BLOCK} and \code{DECAY} are case-insensitive, but the block name, such as \code{UMIX}, is case-sensitive.

        \item \code{Replace}: If none of the aforementioned methods are applicable, users can modify the input file by replacing the value that requires modification with a unique symbol, like \code{ES\_m0} as shown in \reflist{lst1}. Subsequently, during the scan, the unique symbol will be replaced with `m0' using:
\begin{lstlisting}
#               VarID FileID Method Identifier
Input variable: m0, 1, Replace, ES_m0
\end{lstlisting}
         To begin with, \es creates a backup with an additional suffix \code{.ESBackup} in the same directory as the original file. If \es is abruptly interrupted and executed again, the backup file will be automatically employed. When \es successfully completes its operation, the backup file will be deleted.
     \end{itemize}   
    In addition, \es supports a calculation expression and a fixed number as the first element of values. The expression can involve Python's built-in mathematical functions, where commas should be replaced by semicolons. This feature can also be used in \code{Bound} within the \code{[program\#]} section, and in \code{Gaussian} and \code{FreeFormChi2} within the \code{[constraint]} section. For example:
\begin{lstlisting}
Input variable: m0 + m12,        1, Replace,  ES_m0
     	        m0 + pow(m12;2), 1, Position, 2,    2
     	        sqrt(1000),      1, Label,    m0,   2
\end{lstlisting}
         
  \item \code{Output file}: The usage is similar to that of \code{Input file}. Users should note that if a scanning sample cannot generate the desired output file during scanning, that sample will be classified as a non-physical sample.
  
  \item \code{Output variable}: The usage is akin to that of \code{Input variable}, with the distinction that it retrieves values and assigns them to output parameters. This feature supports \code{Position}, \code{Label}, \code{SLHA}, and a new method named \code{File, SAVE}. For example, it can be utilized as follows:
\begin{lstlisting}
#                VarID     FileID  Method
Output variable: spectrum, 1,      File, SAVE
\end{lstlisting}  
  where it will save the file with \code{ID=1}, as defined in \code{Output file}, into a folder named \code{SaveFiles} within the directory specified by \code{Result folder name} in the \code{[scan]} section. The parameter name mentioned here is used in the \code{ScanResult.txt} file, and the parameter value is set to zero. Only files from physical and reserved samples can be stored, with the exception that all files are stored in the \code{OnePoint} sampler. 
  
  If an output variable cannot be assigned a value from its output file, the variable will be set to \code{NaN}, and its associated sample will be marked as non-physical. However, there is one exception: if an output variable cannot be found using the \code{DECAY} method, the variable is set to zero, and its sample is still considered physical. This is because a specific decay may be absent when its branching ratio is zero or very small, as is the case with many spectrum calculators.
  
  \item \code{Bound}: The values can span multiple lines, with each line representing a limit on samples. The first element of each line is the parameter name, and the subsequent elements specify the limits. For example:
\begin{lstlisting}
Bound:  mh,      <=,  130
        mh,      120, 130
        sigmaSI, mn1, max, ../External/limit.txt
\end{lstlisting}
  where the first line represents the condition $\text{mh} \leq 130$, the second line represents the condition $120 < \text{mh} < 130$, and the third line represents the condition that $\text{sigmaSI}$ is less than or equal to an interpolated value based on $mn1$ in the file \code{../External/limit.txt}.
  In the first line, the operator $<=$ can be replaced by $>$, $<$, $>=$, $==$, or $!=$.
  In the third line, the file \code{../External/limit.txt} contains real numbers in two columns, where the values in the first column should be in ascending or descending order. This is used for cases where the limit is on a two-dimensional plane, such as limits on dark matter-nucleon scattering cross-sections in dark matter direct detection experiments. 
  These limits depend on the mass of dark matter. In the example, the first column indicates values of \code{mn1}, and the second column indicates values of \code{sigmaSI}. Based on these columns, \es can obtain a polynomial interpolated value named \code{sigmaSI\_guess} corresponding to \code{mn1}. The operator \code{max} or \code{min} requires that \code{sigmaSI} $\leq$ \code{sigmaSI\_guess} or \code{sigmaSI} $\geq$ \code{sigmaSI\_guess}, respectively. If \code{mn1} is outside the value range of the first column, this bound will be ignored.
    
  \item \code{Time limit minute}: This value should be a positive real number expressed in minutes. It serves to limit the running time of the \code{[program\#]} section for calculating a single sample, ensuring that the running time does not exceed the specified value.

\end{itemize}

\begin{table}[t]
\caption{Rules for the \code{[program\#]} section}
\centering 
\begin{threeparttable}
\begin{tabular}{|c|c|c|c|c|}
\hline
Key                & Type of value                          & \footnotesize{Mandatory} & \footnotesize{Multi-line} & \footnotesize{Component}  \\\hline
\begin{tabular}[c]{@{}c@{}}Execute \\ command \end{tabular}            & string                          & Yes       & Yes        & 1         \\\hline
\begin{tabular}[c]{@{}c@{}}Command \\ path \end{tabular}               & string                          & Yes       & Yes       & 1   \\\hline
\begin{tabular}[c]{@{}c@{}}Input \\ file \end{tabular}   & \begin{tabular}[c]{@{}c@{}}
                        integer,string \\
                     \end{tabular}                          
                                                            & No        & Yes        & 2         \\\hline
\begin{tabular}[c]{@{}c@{}}Input \\ variable \end{tabular}   & \begin{tabular}[c]{@{}c@{}}
                        string,integer,\textbf{Position},integer,integer \\ 
                        string,integer,\textbf{Label},string,integer \\
                        string,integer,\textbf{Replace},string \\
                        string,integer,\textbf{SLHA}, BLOCK, $*$ \\
                        string,integer,\textbf{SLHA}, DECAY, $*$ \\
                     \end{tabular}                              
                                                          & No          & Yes        & $\geq4$         \\\hline
\begin{tabular}[c]{@{}c@{}}Output \\ file \end{tabular}   & integer, string                           & No          & Yes        & 2         \\\hline
\begin{tabular}[c]{@{}c@{}}Output \\ variable \end{tabular}   & \begin{tabular}[c]{@{}c@{}}
                        string,integer,\textbf{Position},integer,integer \\ 
                        string,integer,\textbf{Label},string,integer \\
                        string,integer,\textbf{SLHA}, BLOCK, $*$ \\
                        string,integer,\textbf{SLHA}, DECAY, $*$ \\
                        string,integer,\textbf{File}, SAVE
                     \end{tabular}                      
                                                        & No           & Yes        & $\geq4$         \\\hline
Bound   & string, $*$                                   & No           & Yes        & $\geq3$         \\\hline
\end{tabular}\label{tab:scan2}
\end{threeparttable}
\end{table}

\subsection{\code{[constraint]}}

\begin{table}[ht]
\caption{Rules for the \code{[constraint]} section}
\centering 
\begin{threeparttable}
\begin{tabular}{|c|c|c|c|c|}
\hline
Key                & Type of value                          & Mandatory & Multi-line & Component  \\\hline
Gaussian        & string, float, float, $*$                 & No \tnote{$\dagger$}       & Yes        & $\geq3$          \\\hline
FreeFormChi2            & string                          & No \tnote{$\dagger$}       & Yes        & 1         \\\hline
\end{tabular}\label{tab:scan3}
  \begin{tablenotes}
  \item[$\dagger$] For the \code{MCMC} and \code{MultiNest} samplers, \code{Gaussian} or \code{FreeFormChi2} is mandatory.
  \end{tablenotes}
\end{threeparttable}
\end{table}

\begin{lstlisting} [caption={An example \code{[constraint]} section.}, captionpos=b, label=lst:con]
[constraint]
Gaussian:     mh,   125.0,     2.0
              omg,  0.1199,    0.00012, upper
FreeFormChi2: chisq_from_HiggsSignal
              pow(mh-125;2)/2         
\end{lstlisting}

This section is used to provide guidance for the \code{MCMC} and \code{MultiNest} samplers by defining experimental constraints. Below are the keys and their respective values:
\begin{itemize}
  \item \code{Gaussian}: The \code{Gaussian} values can span across multiple lines, with each line representing a Gaussian distribution. In each line, the first element specifies the parameter name, the second element denotes the central value of the Gaussian distribution, and the third element indicates the standard error at the 95\% confidence level. 
  An optional fourth element can be added, which is either \code{upper} or \code{lower}, indicating upper or lower limits, respectively.
  If this optional element is set to \code{upper} (or \code{lower}), it means that when the parameter value is less (or greater) than the central value, the corresponding $\chi^2 = -2\ln \mathcal{L}$, is set to zero. An example is provided in \reflist{lst:con}.
  
  \item \code{FreeFormChi2}: The \code{FreeFormChi2} values can span across multiple lines, with each line representing an arbitrary expression for the $\chi^2$ value. Each line contains only one element. This method makes it easy to include additional $\chi^2$ contributions calculated in HEP packages, like HiggsSignal, into the total $\chi^2$. An example of this is provided in the last two lines of List~\ref{lst:con}.

\end{itemize}

\subsection{\code{[plot]}}\label{subsec:plot}

\begin{table}[ht]
\caption{Rules for the \code{[plot]} section.}
\centering 
\begin{threeparttable}
\begin{tabular}{|c|c|c|c|c|}
\hline
Key                & Type of value                          & Mandatory & Multi-line & Component  \\\hline
Histogram & string, $*$                                 & No        & Yes        & $\geq1$          \\\hline
Scatter & string, string, $*$                                 & No        & Yes        & $\geq2$          \\\hline
Color & string, string, string, $*$                                 & No        & Yes        & $\geq3$          \\\hline
Contour & string, string, string, $*$                                 & No        & Yes        & $\geq3$          \\\hline
\end{tabular}\label{tab:scan4}
\end{threeparttable}
\end{table}

\begin{lstlisting}[ caption={An example \code{[plot]} section.}, captionpos=b]
[plot]
Histogram: mh
           m1
Scatter:   mn1, sigma
           m0,  m12
Color:     m0,  m12, mh
Contour:   m0,  m12, mh, FigureName
\end{lstlisting}

This section is utilized to create plots for the results generated by \es. Keys and their values are explained as follows:
\begin{itemize}
  \item \code{Histogram}: The values for \code{Histogram} can span across multiple lines, with each line representing a histogram. The first element in a line designates the parameter name, which corresponds to the element within the histogram. The last element specifies the name of the histogram.
  \item \code{Scatter}:
  The values for \code{Scatter} can span across multiple lines, with each line representing a scatter plot. The first and second elements of a line specify the parameter names for the X-axis and Y-axis, respectively. The last element designates the name of the scatter plot. Additionally, for the \code{MCMC} sampler, EasyScan generates a \code{Compare\_FigureName.png} file to compare reserved samples against all physical samples.
  \item \code{Color}: The values are similar to those of \code{Scatter}, with the addition of an element indicating a color bar before the plot name.
  \item \code{Contour}: The values are the same as those of \code{Color}, with the third element indicating a contour line for the graph.
  
\end{itemize}
These figures are stored in the folder named \code{Figures} within the output folder defined by \code{Result folder name} in the \code{[scan]} section.

\subsection{Results from \es}

After scanning is complete, all results will be stored in the folder defined by \code{Result folder name} in the \code{[scan]} section. The folder is located in the same path where \code{easyscan.py} is executed. The contents of the folder may vary depending on the sampler used. Typically, it contains:
\begin{itemize}
	\item \code{ScanResult.txt}: It is used to store numerical results if \es finishes successfully. The output file includes all variables from \code{Input variable} and \code{Output variable} in each \code{[program\#]} section, as well as variables from \code{Gaussian} and \code{FreeFormChi2} in the \code{[constraint]} section. 
    
    Each row in the output file represents one sample, except for the first row, which contains a header indicating the meaning of each column. The header names for variables from the \code{[constraint]} section are \code{Gaussian\_Name} and \code{FreeFormChi2\_Name}. 

    For the \code{MCMC} sampler, two additional columns are included: \code{Chi2}, indicating the total $\chi^2$, and \code{dwell}, indicating the dwell time. In the case of the \code{MultiNest} sampler, two extra columns are \code{Chi2} and \code{probability}.

    For the \code{MultiNest} sampler, only a header is present in \code{ScanResult.txt}, while all the results are stored in a \code{.txt} file located in the folder named \code{MultiNestData}. The columns in the \code{.txt} file correspond to the header in \code{ScanResult.txt}.
    
    To provide an example of \code{ScanResult.txt}, users can refer to \reflist{list:example_result} in \refsec{sec:install}.
	
	\item \code{Figures/}: It is used to save figures defined in the \code{[plot]} section.
	
	\item \code{SavedFile/}: It is used to save output files defined in each \code{[program\#]} section.
	
	\item \code{All\_ScanResult.txt}: It is exclusively employed in the \code{MCMC} sampler, storing all physical samples within it.
\end{itemize}

\section{Examples}\label{sec:example}

The example of a practical application is for investigating the $W$ boson mass and muon $g-2$ anomalies for the Minimal Supersymmetric Standard Model~(MSSM), as presented in Ref.~\cite{Yang:2022gvz}. It can be accomplished by
\begin{lstlisting}
$ cd utils/MSSM_mW
$ make
$ cd ../../
$ ./bin/easyscan.py templates/scan_MSSM_for_mW.ini
\end{lstlisting}
The \code{make} command is used to download and compile \code{SUSYHIT}~\cite{Djouadi:2006bz}, \code{FeynHiggs}~\cite{Heinemeyer:1998yj}, \code{HiggsBounds}~\cite{HiggsBounds}, and \code{GM2Calc}~\cite{Athron:2015rva} separately. It also modifies the input file of \code{SUSYHIT}, \code{suspect2\_lha.in}, to utilize General MSSM at a low scale input. 

\begin{figure}[th]
\centering
\includegraphics[width=0.45\textwidth]{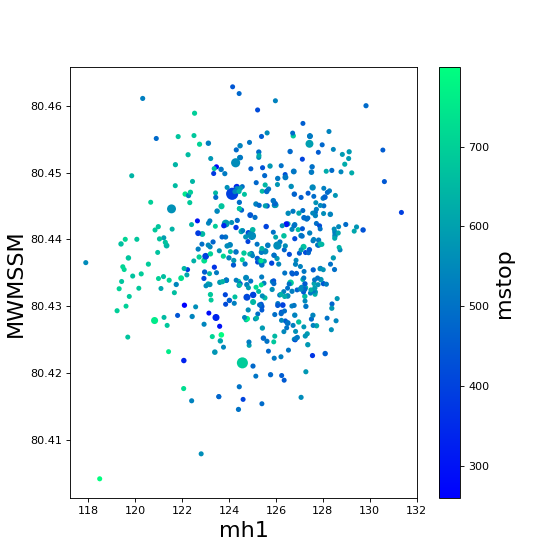}
\includegraphics[width=0.45\textwidth]{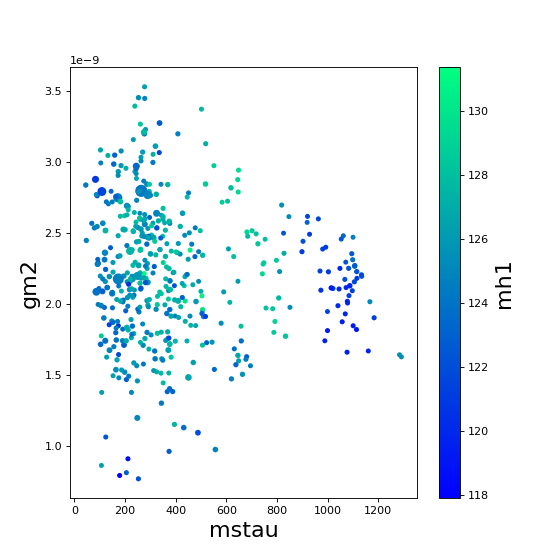}
\caption{Four output plots for the configuration files of \es, \code{example\_random.ini} }
\label{fig:exp1}
\end{figure}

In the configuration file List.~\ref{list:example2}, the \code{MCMC} method is used to generate 20 living samples with 17 free input parameters. These parameters are passed into \code{SUSYHIT} using the SLHA format. Then \code{HiggsBounds} is executed using the output file of \code{SUSYHIT} as an input file, and \code{GM2Calc} is executed using the output file of \code{HiggsBounds} as an input file. The observables include the SM-like Higgs mass from both \code{SUSYHIT} and \code{FeynHiggs}, other masses from \code{FeynHiggs}, $W$ boson mass, and so on. The likelihood is built using $W$ boson mass, $\sin^2 \hat{\theta}$, muon $g-2$, and SM-like Higgs mass.Finally, two plots similar to Fig.~\ref{fig:exp1} will be generated with a smaller number of samples.

\begin{lstlisting}[backgroundcolor=\color{back},frame=trBL,numbers=left,numberstyle=\itshape\color{comment},xleftmargin=2.5em,framexleftmargin=2em,caption={A configuration file for investigating $M_W$ and $\Delta a_\mu$ in the MSSM.}, label={list:example2}]
[scan]
Result folder name: MSSM_MW
Scan method:        mcmc
#                   ID     Prior  Min     MAX     Interval  Initial
Input parameters:   mu,    flat,  100,    2000,   100,       800.6
                    ML123, flat,  100,    2000,   100,       601.7
                    ME123, flat,  100,    2000,   100,       210.2
                    MQ12,  flat,  500,    5000,   100,       4023.7
                    MQ3,   flat,  100,    5000,   100,       763.6
                    MU3,   flat,  100,    5000,   100,       2652.5
                    MD3,   flat,  100,    5000,   100,       4015.1
                    AL,    flat,  -5000,  5000,   100,       3408.8
                    AQ,    flat,  -5000,  5000,   100,       3705.7
                    Ab,    flat,  -5000,  5000,   100,       3158.7
                    At,    flat,  -5000,  5000,   100,       5513.3
                    tanb,  flat,  1,      60,     100,       19.7
                    M3,    flat,  500,    5000,   100,       2204.0
                    MA,    flat,  90,     5000,   100,       1743.5
                    M2,    flat,  100,    1000,   100,       222.0
                    M1,    flat,  100,    1000,   100,       201.1
                    mtop,  flat,  171.86, 173.66, 100,       173.642
Interval of print:  1
Number of points:   20

[program1]
Program name:    susyhit-1.6
Execute command: ./run > susyhit.log
Command path:    utils/susyhit/
Input file:      1,  utils/susyhit/suspect2_lha.in
Input variable:  mtop, 1, SLHA,     BLOCK,  SMINPUTS,   6
                 M1,   1, SLHA,     BLOCK,  EXTPAR,   1
                 M2,   1, SLHA,     BLOCK,  EXTPAR,   2
                 M3,   1, SLHA,     BLOCK,  EXTPAR,   3
                 At,   1, SLHA,     BLOCK,  EXTPAR,   11
                 Ab,   1, SLHA,     BLOCK,  EXTPAR,   12
                 AL,   1, SLHA,     BLOCK,  EXTPAR,   13
                 AQ,   1, SLHA,     BLOCK,  EXTPAR,   14
                 AQ,   1, SLHA,     BLOCK,  EXTPAR,   15
                 AL,   1, SLHA,     BLOCK,  EXTPAR,   16
                 mu,   1, SLHA,     BLOCK,  EXTPAR,   23
                 MA,   1, SLHA,     BLOCK,  EXTPAR,   26
                 tanb, 1, SLHA,     BLOCK,  EXTPAR,   25
                 ML123,1, SLHA,     BLOCK,  EXTPAR,   31
                 ML123,1, SLHA,     BLOCK,  EXTPAR,   32
                 ML123,1, SLHA,     BLOCK,  EXTPAR,   33
                 ME123,1, SLHA,     BLOCK,  EXTPAR,   34
                 ME123,1, SLHA,     BLOCK,  EXTPAR,   35
                 ME123,1, SLHA,     BLOCK,  EXTPAR,   36
                 MQ12, 1, SLHA,     BLOCK,  EXTPAR,   41
                 MQ12, 1, SLHA,     BLOCK,  EXTPAR,   42
                 MQ3,  1, SLHA,     BLOCK,  EXTPAR,   43
                 MQ12, 1, SLHA,     BLOCK,  EXTPAR,   44
                 MQ12, 1, SLHA,     BLOCK,  EXTPAR,   45
                 MU3,  1, SLHA,     BLOCK,  EXTPAR,   46
                 MQ12, 1, SLHA,     BLOCK,  EXTPAR,   47
                 MQ12, 1, SLHA,     BLOCK,  EXTPAR,   48
                 MD3,  1, SLHA,     BLOCK,  EXTPAR,   49

Output file:     1,  utils/susyhit/susyhit_slha.out
Output variable: mh1_hit,  1,  SLHA,  BLOCK,   MASS,    25

[program2]
Program name:    HiggsBounds
Execute command: ./HiggsBounds/build/example_programs/HBSLHAinputblocksfromFH susyhit/susyhit_slha.out > higgsbounds.log
Command path:    utils/

[program3]
Program name:    gm2calc- 2.1.0
Execute command: ./bin/gm2calc.x --slha-input-file=../susyhit/susyhit_slha.out.fh > gm2calc.txt 
Command path:    utils/GM2Calc/
Output file:     1,  utils/GM2Calc/gm2calc.txt
Output variable: mn1,  1,  SLHA,  BLOCK,   MASS,    1000022
                 mn2,  1,  SLHA,  BLOCK,   MASS,    1000023
                 mn3,  1,  SLHA,  BLOCK,   MASS,    1000025
                 mn4,  1,  SLHA,  BLOCK,   MASS,    1000035
                 mc1,  1,  SLHA,  BLOCK,   MASS,    1000024
                 mc2,  1,  SLHA,  BLOCK,   MASS,    1000037
                 mh1,  1,  SLHA,  BLOCK,   MASS,    25
                 mh2,  1,  SLHA,  BLOCK,   MASS,    35
                 mstau,1,  SLHA,  BLOCK,   MASS,    1000015
                 mstop,1,  SLHA,  BLOCK,   MASS,    1000006
                 msb,  1,  SLHA,  BLOCK,   MASS,    1000005
                 MWMSSM,     1, SLHA,  BLOCK,   PRECOBS,    3
                 MWSM,       1, SLHA,  BLOCK,   PRECOBS,    4
                 SW2effMSSM, 1, SLHA,  BLOCK,   PRECOBS,    5
                 SW2effSM,   1, SLHA,  BLOCK,   PRECOBS,    6
                 gm2,        1, SLHA,  BLOCK,   GM2CalcOutput,    0

[constraint]
#           varID       mean       uncertainty
Gaussian:   MWMSSM,     80.452,    0.013453624
            SW2effMSSM, 0.23121,   0.00010770330
            gm2,        25.1E-10,  6.2297673E-10
            mh1,        125.4,     3.
            
[plot]
color: mh1, MWMSSM, mstop
       mstau, gm2, mh1
\end{lstlisting}

\section{Summary}
\label{sec:summary}
This paper introduces the workflow and usage of \es. Essentially, \es provides an easy-to-use platform with a rich input and output interface for connecting different HEP packages. Then we can easily use the built-in scanning algorithms to study the parameter space of different models. It also provides an automated visualization module that can quickly understand the results at the end of the scan.
To keep track of developments, report bugs or ask for help, please see  \url{https://github.com/phyzhangyang/EasyScan_HEP}.

\section{Acknowledgments}
This work was supported by the National Natural Science Foundation of China (NNSFC) under grant Nos. 12105248, 12047503, 11705048, China Scholarship Council No.202208410277, and by Peng-Huan-Wu Theoretical Physics Innovation Center.

\bibliographystyle{utphys}
\bibliography{cit}

\end{document}